\documentclass[multphys,vecphys]{svmult}
\usepackage{makeidx}         
\usepackage{graphicx}        
\usepackage{multicol}        
\usepackage{cite}            
\usepackage[bottom]{footmisc}

\makeindex             

\begin{document}

\title{Microwave spectroscopy of Q1D and Q2D organic conductors}
\titlerunning{Microwave spectroscopy of organic conductors}
\author{Stephen Hill \and Susumu Takahashi}
\institute{Department of Physics, University of Florida,
Gainesville, FL32611-8440, USA \texttt{hill@phys.ufl.edu}}

\maketitle

This chapter reviews recent experimental studies of a novel
open-orbit magnetic resonance phenomenon. The technique involves
measurement of angle-dependent microwave magneto-conductivity and
is, thus, closely related to the cyclotron resonance and
angle-dependent magnetoresistance techniques. Data for three
contrasting materials are presented: (TMTSF)$_2$ClO$_4$,
$\alpha$-(BEDT-TTF)$_2$KHg(SCN)$_4$ and
$\kappa$-(BEDT-TTF)$_2$I$_3$. These studies reveal important
insights into the Fermiology of these novel materials, as well as
providing access to important electronic parameters such as the
in-plane Fermi velocity and quasiparticle scattering rate. It is
argued that all three compounds exhibit coherent three-dimensional
band transport at liquid helium temperatures, and that their
low-energy magnetoelectrodynamic properties appear to be well
explained on the basis of a conventional semiclassical Boltzmann
approach. It is also suggested that this technique could be used to
probe quasiparticles in nodal superconductors.

\section{Introduction}
\label{sec:1}

Many of the novel broken symmetry states observed in organic
conductors are driven by electronic instabilities associated with
their low-dimensional Fermi surfaces (FSs), e.g. nesting
instabilities \cite{nesting,KagBook,GrunerB,IYS,Louati,Tanuma}.
Consequently, techniques which can probe the detailed topology of
the FSs of organic charge transfer salts have been widely employed
by researchers in this field (for several recent reviews, see
\cite{Wosnitza,JSReview,JSBook,KartsReview}). Examples include: the
de Haas-van Alphen (dHvA) and Shubnikov-de Haas (SdH) effects
\cite{Wosnitza,JSReview,JSBook,KartsReview,Shoen,Abrikos};
angle-dependent magnetoresistance oscillations (AMRO)
\cite{Wosnitza,JSReview,JSBook,KartsReview,Lebed86,LebedBak,gsbPRL90,OsadaPRL91,0onPRL91,OsadaPRL92,Karts93,DKC,BlundPRB96,Chash97,Chash98,Lee98,KangSM03,LebedPRL04,YakovPRL06};
angle-resolved photoemission spectroscopy (ARPES) \cite{Zwick97};
and cyclotron resonance (CR)
\cite{Perel91,JSPRL92,HillSM93,HillSM95,HillPRB96,Poliskii,HillPRB97,BlundPRB97,OhtaSM97a,OhtaSM97b,OhtaSM97c,HillPB98,ArdPRL98,ArdPRB99,Palassis,OhtaSM99a,OhtaSM99b,OhtaSM01,KovalevPRB02,KovalevJAP03,KovalevPRL03,OhtaJPSJ02,OhtaJPSJ03,OhtaPRB03,SusumuIJMP,SusumuJAP05,SusumuPRB05}.
With the exception of ARPES, all of the above techniques involve the
use of strong magnetic fields and low-temperatures. In the case of
the SdH and dHvA effects, the magnetic field causes Landau
quantization which leads to the magneto-oscillatory behavior of
various thermodynamic and transport phenomena \cite{JSBook,Shoen}.
The CR and AMRO effects, meanwhile, are essentially semiclassical in
origin, and are caused by the periodic motion of electrons induced
by the Lorentz force
\cite{OsadaPRL92,BlundPRB96,HillPRB97,BlundPRB97}. Each of these
techniques requires that the product $\omega_c\tau > 1$, where
$\omega_c$ ($=eB/m^*$) is the cyclotron frequency and $\tau$ is the
transport scattering time. Meanwhile, the SdH and dHvA effects
additionally require that $\hbar\omega_c > \{k_B T , h /
\tau_\varphi\}$, where $\tau_\varphi$ is the quantum lifetime
\cite{Shoen}. These criteria are easily satisfied for many organic
conductors due to their exceptional purity. However, instances of
the application of such methods to inorganic oxides such as the
cuprate and ruthenate superconductors are extremely rare
\cite{MacKenzieRMP,HillPRL00,Hussey}. Consequently, ARPES is
probably the technique which has been most widely applied, and has
contributed most significantly to the understanding of FS driven
phenomena in unconventional (including high-T$_c$) superconductors
\cite{ARPES}.

In this chapter, we describe a new open-orbit magnetic resonance
phenomenon, the so-called periodic-orbit resonance (POR), which
enables angle-resolved mapping of the in-plane Fermi velocity
($v_F$) for both quasi-one-dimensional (Q1D) and
quasi-two-dimensional (Q2D) organic conductors
\cite{HillPRB97,BlundPRB97,KovalevPRB02,KovalevJAP03,KovalevPRL03}.
As such, this technique is complimentary to ARPES, i.e. it can
provide information concerning the in-plane momentum dependence of
the density-of-states ($\propto v_F$) and quasiparticle scattering
rate ($\tau^{-1}$). However, the POR phenomenon involves measurement
of the bulk microwave conductivity \cite{HillPRB00}. Consequently,
it is immune to surface effects which have been known to cause
problems in ARPES measurements. Furthermore, the POR technique
provides sub-millivolt energy resolution and is, thus, sensitive to
extremely fine details of the FS topology
\cite{HillPRB97,BlundPRB97,KovalevPRB02,KovalevPRL03,SusumuPRB05}.
We will illustrate the utility of this method for several organic
conductors, including (TMTSF)$_2$ClO$_4$ and (BEDT-TTF)$_2$X [X =
KHg(SCN)$_4$ and I$_3$].

We begin by developing a theoretical framework which enables us to
simulate the microwave magneto-conductance of a sample with a Q1D FS
topology (see Fig.~\ref{fig:FS}). We follow exactly the same
semiclassical approach which has been adopted by various researchers
to model AMRO data \cite{OsadaPRL92,BlundPRB96}. Indeed, POR
represent nothing more than an evolution of AMRO to high
frequencies, such that $\omega \sim \omega_c$ and $\omega_c\tau
> 1$, where $\omega$ is the measurement frequency and $\omega_c$
represents the characteristic frequency associated with any periodic
motion of quasiparticles on the FS induced by the Lorentz force.
AMRO occur due to the commensurate motion of electrons on the FS for
certain applied field orientations (provided $\omega_c\tau > 1$).
These commensurabilities affect the collective dynamical properties
of the electronic system, giving rise to resistance minima
(`resonances') when the field is aligned with real space lattice
vectors. The idea of ``magic angle'' resonances goes back 20 years
to Lebed \cite{Lebed86}, who first predicted the occurrence of
angle-dependent magnetic field effects in organic conductors.
Lebed's theory predicted dimensional crossovers
(3D$\rightarrow$2D$\rightarrow$1D) arising from commensurate
electronic motion \cite{Lebed86,LebedBak,LebedPRL04}. Some authors
have since suggested that this could lead to Fermi liquid/non-Fermi
liquid (FL/NFL) crossovers \cite{CSCC}, i.e. to fundamental
differences in the thermodynamic ground state at, and away from,
Lebed's ``magic angles''.

When one moves to a rotating frame corresponding to incident
microwave radiation of frequency $\omega\approx\omega_c$, one finds
that the semiclassical commensurability effects discussed above
occur at field orientations which depend on the frequency
\cite{BlundPRB97,KovalevPRB02,SusumuPRB05,YakovPRL06}. In other
words, there is nothing ``magic'' about the angles corresponding to
the AMRO minima, as we will demonstrate from POR (AC AMRO)
measurements presented in this chapter. Experiments also suggest
that, at temperatures below 5~K and at moderate fields, the
quasiparticle dynamics for all of the above mentioned compounds
appear to be coherent in all three directions
\cite{KovalevPRB02,KovalevPRL03,SusumuPRB05}. Indeed, the presented
results are consistent with a 3D Fermi-liquid state, and can be
easily interpreted in terms of the semiclassical Boltzmann transport
equation (or an equivalent quantum mechanical theory
\cite{ArdPRB99,YakovPRL06}). These findings are somewhat at odds
with recent thermal transport measurements
\cite{WuPRL93,WuPRL05,ChoiPRL05} presented elsewhere in this review,
and do not support the idea of FL/NFL crossovers. Finally, we
consider open-orbit POR in Q2D systems subjected to an in-plane
magnetic field \cite{KovalevPRL03}, and discuss the possibility of
utilizing this technique to probe the normal quasiparticles in nodal
superconductors \cite{SusumuJAP05}.

\section{The Periodic Orbit Resonance Phenomenon}
\label{sec:2}

\begin{figure}[t]
\centering
\includegraphics*[width=0.95\columnwidth]{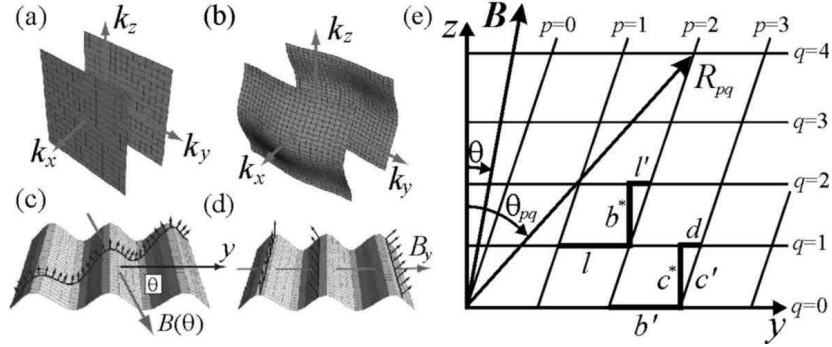}
\caption[]{Fermi surfaces corresponding to Eq.~\ref{eqn:1} with (a)
$t_b=t_c=0$, and (b) finite $t_b$ and $t_c$, which causes a warping
of the flat 1D FS in (a). The trajectories of quasiparticles on a
warped FS (finite $t_b$, and $t_c=0$) are shown in (c) and (d) for
different field orientations; the arrows represent the quasiparticle
velocities. In (c), the field results in an oscillatory $v_y$,
whereas this is not the case in (d), where the field is applied
along the FS warping direction. (e) The oblique real-space lattice
appropriate to the tight-binding model represented by
Eq.~\ref{eqn:5}. The relevant real-space vectors $\vec{R}_{pq} =
(0,pb^\prime+qd,qc^*)$ for (TMTSF)$_2$ClO$_4$, and
$(0,pl+ql^\prime,qb^*)$ for $\alpha$-(BEDT-TTF)$_2$KHg(NCS)$_4$
(section~\ref{sec:6}).} \label{fig:FS}
\end{figure}

We shall mainly limit discussion here to open-orbit POR; for
detailed discussion of closed orbit POR observed in Q2D systems,
refer to \cite{HillPRB97,BlundPRB97}. As a starting point, we
consider a Q1D system with a pair of corrugated FS sheets at $k_x =
\pm k_F$ (see Fig.~\ref{fig:FS}(b)). Such a FS is typical for the
Q1D Bechgaard salts, where electrons delocalize easily along the
TMTSF cation chains due to linear stacking of the partially occupied
Se $\pi$-orbitals \cite{IYS,KagBook}. The FS corrugations arise from
the weak orbital overlap transverse to the chain direction. Within a
simple tight binding scheme, such an electronic band structure may
be parameterized in terms of a set of highly anisotropic transfer
integrals $(t_a:t_b:t_c\approx 200:20:1$~meV for (TMTSF)$_2$ClO$_4$
\cite{IYS}). We begin a simple treatment of the POR phenomenon by
considering an orthorhombic crystal structure, and by linearizing
the $x$-axis dispersion about $\epsilon_F~(=0)$ and $k_x = \pm k_F$
(we treat a more general case below). The energy dispersion is then
expressed as,

\begin{equation}\label{eqn:1}
E({\vec{k}}) = \hbar v_F (|k_x | - k_F ) - 2t_b \cos (k_y b) - 2t_c
\cos (k_z c),
\end{equation}

\noindent{where $b$ and $c$ are the $y$ and $z$ dimensions of the
orthorhombic unit cell.} The simplest case involves setting
$t_b=t_c=0$, in which case the FS consists of a pair of absolutely
flat sheets at $k_x = \pm k_F$ (Fig.~\ref{fig:FS}(a)). A finite
$t_b$ results in a sinusoidal corrugation of the FS, with
periodicity $2\pi/b$ directed along the $b$- (or $y$-) axis. This
situation is illustrated in Figs.~\ref{fig:FS}(c) and (d).

We next consider the affect of the Lorentz force [$\hbar
\dot{\vec{k}}=-e(\vec{v}_F\times \vec{B})$] on a quasiparticle on
the FS due to a magnetic field, $\vec{B}$, applied within the
$yz$-plane (i.e. parallel to the FS). The quasiparticle follows a
trajectory over the corrugated FS which is perpendicular to
$\vec{B}$ (Fig.~\ref{fig:FS}(c)). For arbitrary orientation of the
field within the $yz$-plane, this gives rise to motion which is
periodic with a characteristic frequency

\begin{equation}\label{eqn:2}
\omega _c  = 2\pi \frac{{\dot k_y }}{{2\pi /b}} = \frac{{v_F
eBb}}{\hbar }|\sin \theta |,
\end{equation}

\noindent{where $\theta$ is the angle between the magnetic field and
the corrugation axis ($b$-axis in this case)}. This periodic
$k$-space motion has important consequences for the conductivity
which, in the Boltzmann picture, is governed by the time evolution
of quasiparticle velocities averaged over the FS, i.e.

\begin{equation}\label{eqn:3}
\sigma _{ii}(\omega)  \propto -\int (\frac{{\partial f_{\vec{k}}
}}{{\partial \varepsilon _{\vec{k}} }}{\rm{)}}{v}_i
({\vec{k}},0){d^3 {\vec{k}}\int\limits_{ - \infty }^0 {v_i
({\vec{k}},t)} } e^{ - i\omega t} e^{t/\tau }dt,
\end{equation}

\noindent{where $i= x, y$ or $z$.} Naturally, the time evolution of
each velocity component is governed by the field strength and its
orientation. For small corrugation ($t_b\ll t_a$), one can neglect
variations in $v_x$ to the first order of approximation. However, it
is clear that $v_y$ will acquire an oscillatory component with a
frequency $\omega_c$ given by Eq.~\ref{eqn:2}. Inserting this time
dependence into Eq.~\ref{eqn:3} results in the following expression
for the $y$-component of the conductivity tensor:

\begin{equation}\label{eqn:4}
{\mathop{\rm Re}\nolimits}\{\sigma _{yy}(\omega,B,\theta)\}  =
\frac{{\sigma _\circ }}{2}\left[ {\frac{1}{{1 + (\omega  + \omega _c
)^2 \tau ^2 }} + \frac{1}{{1 + (\omega  - \omega _c )^2 \tau ^2 }}}
\right],
\end{equation}

\noindent{where $\sigma_\circ$ is the DC conductivity. This equation
reduces to the simple Drude formula for the AC conductivity of a
metal for $B=0$ (i.e. $\omega_c=0$).} For finite magnetic field,
Eq.~\ref{eqn:4} contains non-resonant and resonant terms where the
$\omega$ in the denominator of the Drude formula is replaced by
$(\omega + \omega_c)$ and $(\omega - \omega_c)$, respectively.

It is important to note that $\omega$ and $\omega_c$ essentially
play identical roles in Eq.~\ref{eqn:4}, hence the correspondence
between the AMRO and POR phenomena. In a finite magnetic field, the
DC conductivity $\sigma_{yy}(\omega=0,\theta)$ exhibits a resonance
(resistance minimum) when $\omega=\omega_c=0$, i.e. when the
magnetic field is directed along the $b$-direction corresponding to
a ``magic angle" ($\theta=0$). From Fig.~\ref{fig:FS}(d), it is
clear that when the applied field is directed along the corrugation
axis (magic angle), it has no influence on either the $y$ or $z$
components of the quasiparticle velocities. Consequently, the usual
Drude behavior is recovered, and both $\sigma_{yy}$ and
$\sigma_{zz}$ exhibit a maximum at $\omega=0$. As the field is
rotated away from the magic angle, $\omega_c$ becomes finite and
increases as $\sin\theta$. Consequently, $\sigma_{yy}(\omega=0)$
decreases ($\rho_{yy}$ increases); the finite $\omega_c$ essentially
leads to a randomization of $v_y$ between successive collisions and,
therefore, to a suppression of $\sigma_{yy}$. This is the origin of
the resistance minima observed in AMRO experiments (see further
discussion below). For a finite $\omega_c$ ($\theta \neq 0$),
meanwhile, the resonance can be recovered by setting
$\omega=\omega_c$; in this situation, $v_y$ remains static in a
rotating frame corresponding to the driving frequency $\omega$. This
is the origin of the POR phenomenon \cite{HillPRB97,BlundPRB97},
which was originally discussed by Osada et al. \cite{OsadaPRL92}.
What one sees, therefore, is that AMRO correspond to DC conductivity
resonances observed by sweeping $\omega_c$, while POR correspond to
precisely the same resonances except that they are observed at
finite $\omega$, i.e. the Drude conductivity peak moves away from
$\omega=0$ to $\omega=\omega_c$. As we shall demonstrate, POR are
also best observed by sweeping $\omega_c$; this may be achieved
either by varying $B$ directly, or by rotating the field in exactly
the same way as one would in an AMRO experiment. For the field
rotation measurements (AC AMRO), we note that two resonances should
be observed at $\theta=\pm\sin^{-1}\{\hbar\omega/v_F e Bb\}$, not a
single resonance at $\theta=0$. Thus, from this semiclassical
point-of-view, one clearly sees that the directions corresponding to
AMRO minima are not really ``magic".

The preceding discussion assumes the existence of an extended 3D FS,
i.e. that the sample under investigation exhibits coherent 3D band
transport. Given the extreme anisotropy of many organic conductors,
it is natural to consider what would happen if the quasiparticle
dynamics were incoherent in one or more directions. A number of
authors have explored this limit in which the scattering rate
exceeds the hopping frequency in a given direction within a crystal
\cite{Moses1,Moses2,Osadatc}, e.g. $\tau^{-1}>t_c/\hbar$. In this
case, it is meaningless to consider energy dispersion and FS warping
in this direction; essentially, the lifetime broadening of
quasiparticle states exceeds the bandwidth in that direction.
McKenzie et al. have shown that, for layered materials, POR occur
for both coherent and weakly incoherent interlayer transport
\cite{Moses2}. Thus, the observation of POR does not necessarily
imply the existence of a 3D FS. By weakly incoherent, it is implied
that one cannot define an interlayer momentum ($\hbar k_z$), but
that the in-plane momentum is conserved when a quasiparticle hops
between layers \cite{Moses1,Moses2}. Therefore, it is likely that
many of the conclusions drawn from the POR studies described in this
chapter apply regardless of whether a truly 3D FS exists.
Nevertheless, we will argue that all of the materials investigated
in this chapter exhibit a coherent 3D Fermi liquid state at low
temperatures ($< 5$~K).

In principle, one could also observe POR by sweeping $\omega$, just
as one can use zero-field optical techniques to measure the Drude
conductivity in a conventional metal \cite{DressGrunBook}. However,
the large effective masses ($\sim 2 m_e$) and exceptionally long
scattering times (tens of ps) found in many organic conductors at
liquid helium temperatures make it very difficult to observe POR
using broadband techniques, due to lack of sensitivity and/or
resolution at microwave frequencies \cite{DressPRL96}. Hence the
need for high-sensitivity narrow-band cavity-based techniques
\cite{Red1,Red2,Red3,MolaRSI,SusumuRSI}. As an aside, as noted
above, frequency- and field-swept experiments are essentially
equivalent in the Boltzmann theory. Consequently, one should be able
to extract essentially the same information from a POR measurement
as one would from a broadband optical measurement. We will make such
comparisons in section~\ref{sec:11} of this chapter.

\subsection{Modification of theory for realistic crystal structures}
\label{sec:3}

As shown in the previous section, a finite tight-binding transfer
integral along the $y$-direction (perpendicular to the highly
conducting $x$-axis) results in a single resonance in the
$y$-component of the conductivity tensor ($\sigma_{yy}$) when
$\omega=\omega_c$. In the preceding example, we set $t_c=0$.
Consequently, $v_z=0$ for all momenta, and it is trivial to show
that this leads to $\sigma_{zz}=0$. If $t_c$ were finite, the FS
would acquire an additional corrugation along the orthogonal
$z$-direction. For such an orthorhombic crystal, one can show that
the Lorentz force leads to completely separable solutions for
$v_y(t)$ and $v_z(t)$ (because the corrugation axes are orthogonal).
In fact, $v_y(t)$ [$=2bt_b\sin\{k_y(t)b\}/\hbar$] depends only on
$t_b$ and $v_z(t)$ [$=2ct_c\sin\{k_z(t)c\}/\hbar$] depends only on
$t_c$. Now that $v_z$ also oscillates, one would expect to observe a
DC AMRO minimum in $\sigma_{zz}$ when the field is directed exactly
along the $c$-axis. At finite frequencies, this resonance would
evolve into two POR at angles
$\theta^\prime=\pm\sin^{-1}\{\hbar\omega/v_F e Bc\}$ where, this
time, $\theta^\prime$ is the angle between the applied field and the
$c$- (or $z$-) axis. The important point to note here is that only
$\sigma_{zz}$ is sensitive to $t_c$ and only $\sigma_{yy}$ is
sensitive to $t_b$. Thus, one would need to independently measure
the conductivities along orthogonal directions in such a crystal in
order to extract information about the two transverse tight-binding
transfer integrals. We note that an absolute measure of these
parameters would require absolute measurements of the conductivity,
which is extremely difficult using a narrow-band cavity perturbation
technique \cite{DressPRL96,Red1,Red2,Red3}. However, as we shall now
see, one can directly obtain information concerning the relative
magnitudes of different transfer integrals from measurement of a
single conductivity component for crystal structures with symmetry
lower than orthorhombic \cite{BlundPRB96,KovalevPRB02}.

We now consider the quite general case of an oblique lattice
\cite{BlundPRB96,KovalevPRB02}, as illustrated in
Fig.~\ref{fig:FS}(e). We also consider the possibility of electron
hopping not only between nearest-neighbor chains, but also to
next-nearest-neighbors and so on. In the standard fashion, we label
chains by a pair of indices $p$ and $q$, as illustrated in
Fig.~\ref{fig:FS}(e). For convenience, we treat the specific case of
the (TMTSF)$_2$ClO$_4$ lattice for which the crystallographic
$a$-axis corresponds to the chain direction, $b^\prime$ ($\perp a$,
$\| ab$-plane) corresponds to the intermediate conducting direction,
and $c^*$ ($\perp ab$-plane) corresponds to the least conducting
direction \cite{IYS}. We then reference this lattice to a Cartesian
coordinate system by aligning the crystallographic $a$ and
$b^\prime$-directions with the $x$- and $y$-axes, respectively. The
real-space vectors $\vec{R}_{pq}$ which define the locations of
neighboring chains are then given by $(0,pb^\prime+qd,qc^*)$, where
$p$ and $q$ are integers and $d$ is defined in Fig.~\ref{fig:FS}(e).
One can then write down a linearized dispersion relation about
$\epsilon_F~(=0)$ and $k_x=\pm k_F$ in terms of a set of
tight-binding transfer integrals, $t_{pq}$, between neighboring
chains, i.e.

\begin{equation}\label{eqn:5}
\begin{array}{l}
E(\vec k) = \hbar v_F (|k_x | - k_F ) - \sum\limits_{p,q} {t_{pq}
\cos (\vec{k}\cdot \vec{R}_{pq})} \\ \\ \ \ \ \ \ \ \ = \hbar v_F
(|k_x | - k_F ) - \sum\limits_{p,q} {t_{pq} \cos [(pb'
+ qd)k_y  + (qc^* )k_z ]}.\\
\end{array}
\end{equation}

\noindent{Each transfer integral $t_{pq}$} will produce a sinusoidal
corrugation of the FS, with the corrugation axis defined by the
real-space vector $\vec{R}_{pq}$, and the period given by
$2\pi/|\vec{R}_{pq}|$; the amplitudes of the corrugations will scale
with $t_{pq}$. More importantly, each $t_{pq}$ will produce a
sinusoidal modulation of the quasiparticle velocity components
transverse to the chains, and to a corresponding resonance in the
conductivity \cite{BlundPRB96}. However, the dispersion relation
given by Eq.~\ref{eqn:5} does not give rise to completely separable
solutions for $v_y(t)$ and $v_z(t)$. In particular, hopping in the
$c$-direction (finite $q$) will have a direct influence on $v_y(t)$.
Furthermore, diagonal hopping terms involving finite $p$ and $q$
will affect both velocity components and, consequently, give rise to
resonances in both $\sigma_{yy}$ and $\sigma_{zz}$. Thus, in
general, one may now expect to observe several POR harmonics (or
AMRO minima) depending on the relative magnitudes of the higher
order $t_{pq}$ transfer integrals.

In order to parameterize this more general case, one can define a
set of angles, $\theta_{pq}$, corresponding to interchain directions
[see Fig.~\ref{fig:FS}(e)]. Then,

\begin{equation}\label{eqn:6}
\tan \theta _{pq}  = \frac{{pb'}}{{qc^* }} + \frac{d}{c^*},
\end{equation}

\noindent{where this angle is referenced to the $z$-direction
[$c^*$-axis for (TMTSF)$_2$ClO$_4$] within the plane of the FS
($yz$-plane)} \cite{KovalevPRB02}. These are the so-called Lebed
magic angles \cite{Lebed86}. One can then go through the same
procedure as in the preceding section in order to consider the
effect of an applied magnetic field. For field rotation in the plane
of the FS, the characteristic POR frequencies are given by

\begin{equation}\label{eqn:7}
\omega_{pq}  = \frac{{v_F eBR_{pq} }}{\hbar }
|\sin(\theta-\theta_{pq})| = \omega_{pq}^{max}
|\sin(\theta-\theta_{pq})|,
\end{equation}

\noindent{where $\theta$ is again measured relative to the $z$-axis,
and $\omega_{pq}^{max}~(=v_F e B R_{pq}/\hbar)$ is the maximum value
of $\omega_{pq}$ occurring when $|\theta - \theta_{pq}| = 90^\circ$
\cite{KovalevPRB02}. The transverse components of the conductivity
tensor ($\sigma_{yy}$ and $\sigma_{zz}$) may then be written}

\begin{equation}\label{eqn:8}
{\mathop{\rm Re}\nolimits} \{\sigma _{ii} (\omega ,\theta ,B)\}
\propto \sum\limits_{p,q} {A_{ii}^{pq} \left[\frac{1}{{1 + (\omega +
\omega _{pq} )^2 \tau ^2 }} + \frac{1}{{1 + (\omega  - \omega _{pq}
)^2 \tau ^2 }}\right]},
\end{equation}

\noindent{where $A_{yy}^{pq} = \{(pb^\prime + qd)t_{pq}\}^2$} and
$A_{zz}^{pq} = \{qc^*t_{pq}\}^2$. From these expressions, one can
see that DC AMRO minima occur when $\theta=\theta_{pq}$
($\omega_{pq}=0$), provided the appropriate $A_{yy}^{pq}$ is finite.
However, in general, POR will be observed when $\omega =
\omega_{pq}$, at angles given by Eq.~\ref{eqn:7}, i.e.

\begin{equation}\label{eqn:9}
\theta_{POR} = \theta_{pq} \pm \sin^{-1}(\omega/\omega_{pq}^{max}).
\end{equation}

\noindent{Here, one again sees that the resonance condition reduces
to the DC result, $\theta = \theta_{pq}$, when $\omega = 0$.

Each resonance, be it a DC AMRO minimum or a POR, corresponds to a
given transfer integral, $t_{pq}$. As noted above, the amplitude of
each POR is proportional to $A_{ii}^{pq}\propto t_{pq}^2$. Thus, the
relative intensities of POR provide a direct measure of the fourier
spectrum of the FS warping, since the $t_{pq}$ represent the fourier
amplitudes associated with each modulation vector $\vec{R}_{pq}$
\cite{BlundPRB96}. We note that DC AMRO minima for a given $p/q$
occur at the same angles, irrespective of the values of $p$ and $q$.
This is not the case for a finite frequency POR measurement, where
the resonance depends on both $\theta_{pq}$ and $R_{pq}$ through the
prefactor in Eq.~\ref{eqn:7}. Indeed, the POR condition
(Eq.~\ref{eqn:7}) defines a set of curves in the 2D $\omega$ versus
$\theta$ plane, whereas AMRO correspond simply to the 1D line of
points representing to the $\omega=0$ intercepts of the POR curves.
The POR data presented in the following section have been performed
at many different frequencies, both by sweeping the field
(equivalent to sweeping $\omega$) at a fixed orientation relative to
the crystal, and by varying the orientation (sweeping $\theta$) of a
static field.

\section{Experimental observation of POR for Q1D systems}
\label{sec:4}

In this section, we present experimental POR data obtained for two
contrasting Q1D systems: (TMTSF)$_2$ClO$_4$ and
$\alpha$-(BEDT-TTF)$_2$KHg(SCN)$_4$. Microwave measurements were
carried out using a millimeter-wave vector network analyzer and a
high sensitivity cavity perturbation technique; this instrumentation
is described in detail elsewhere \cite{MolaRSI,SusumuRSI}. In order
to enable in-situ rotation of the sample relative to the applied
magnetic field, two different techniques were employed. The first
involved a split-pair magnet with a 7~T horizontal field and a
vertical access. Smooth rotation of the entire rigid microwave probe
relative to the fixed field was achieved via a room temperature
stepper motor (with $0.1^\circ$ resolution). The second method
involved in-situ rotation of the end-plate of a cylindrical cavity,
mounted with its axis transverse to a 17~T superconducting solenoid.
Details concerning this cavity, which provides an angle resolution
of $0.18^\circ$, have been published elsewhere \cite{SusumuRSI}; a
combination of the two methods enables double-axis rotation
capabilities.

\subsection{(TMTSF)$_2$ClO$_4$}\label{sec:5}

(TMTSF)$_2$ClO$_4$ undergoes a structural transition at
$T_{AO}=24$~K which involves an ordering of the ClO$_4$ anions
(which lack a center of inversion) \cite{IYS}. The anion order,
which results in a doubling of the unit cell along the
$b$-direction, is extremely sensitive to the cooling rate around
$T_{AO}$. When cooled rapidly (quenched), at rates exceeding
50~K/min, the ClO$_4$ anions remain disordered and the system
undergoes a transition to a spin-density-wave (SDW) insulating
ground state below 6~K. On the other hand, if the sample is cooled
slowly (relaxed) through $T_{AO}$ at a rate of less than 0.1~K/min,
the anions order. Relaxed samples remain metallic and eventually
become superconducting below about 1~K. Thus, great care was taken
in all experiments on (TMTSF)$_2$ClO$_4$ to ensure reproducible
cooling of the sample. Indeed, all of the data presented in this
chapter were obtained for samples cooled at rates between 0.01 and
0.1~K/min from 32~K to 17~K. Even for these slow cooling rates,
subtle differences are found in the data for different cooling
rates.

Figure~\ref{fig:ClO4data} shows changes in the microwave loss
measured in a cylindrical TE011 ($f_{011}=52.1$~GHz) cavity
containing a single (TMTSF)$_2$ClO$_4$ crystal at a temperature of
2.0~K; the frequency is 52.1~GHz in (a) and 75.5~GHz in (b) and (c).
The sample was positioned so that the $c^*$-axis conductivity
dominates the losses in the cavity arising from the perturbation due
to the sample \cite{HillPRB00}. The $c^*$-axis conductivity is
sufficiently low and the sample sufficiently thin so that
microwave-induced $c^*$-axis currents penetrate more-or-less
uniformly throughout the sample (depolarization regime). In this
regime, the loss arising due to the sample is proportional to the
conductivity \cite{Red1,Red2,Red3}. Therefore, one can equivalently
consider the vertical axis in Fig.~\ref{fig:ClO4data} to represent
AC conductivity. In Fig.~\ref{fig:ClO4data}(a), data are plotted
(offset) as a function of the orientation of the applied magnetic
field within the $b^\prime c^*$-plane, for several applied field
strengths. In Figs.~\ref{fig:ClO4data}(b) and (c), data are plotted
(offset) as a function of the field strength, for many field
orientations.

\begin{figure}[t]
\centering
\includegraphics*[width=1\columnwidth]{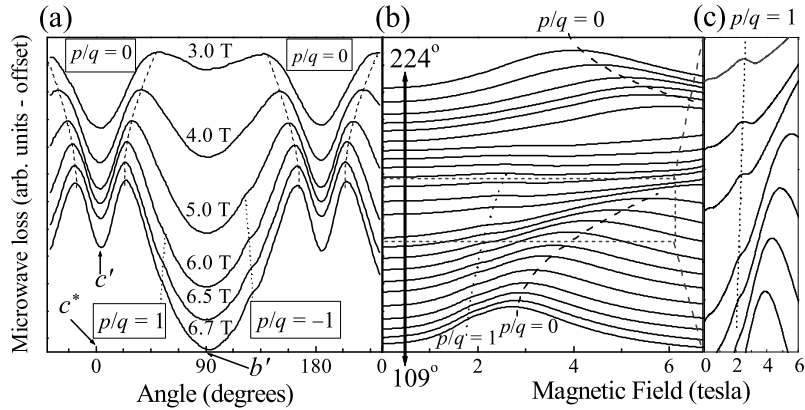}
\caption[]{Angle- (a) and field-swept [(b) and (c)] microwave
absorption data ($\propto\sigma_{zz}$) for (TMTSF)$_2$ClO$_4$
(sample C) \cite{SusumuPRB05}. The peaks correspond to POR (offset
for clarity), which have been labeled according to the scheme
described in the main text. The field was rotated in the
$b'c^*$-plane; the temperature was 2~K, and the frequency was
52.1~GHz in (a) and 75.5~GHz in (b) and (c). The angle step in (b)
is $5^{\circ}$, and the data in (c) correspond to an enlargement of
the small resonances denoted as $p/q=1$ in (b).}
\label{fig:ClO4data}
\end{figure}

The first thing to note is the resemblance of the data in
Fig.~\ref{fig:ClO4data}(a) to DC AMRO data, albeit inverted
(conductivity rather than resistivity is plotted). Indeed, the
quality of the data are comparable to the best AMRO measurements
\cite{KangSM03}, even though the present technique is contactless
and extremely sensitive to the mechanical stability of the
instrumentation. Clear conductivity resonances (resistance minima)
are seen either side of the positions close to $0^\circ$ and
$180^\circ$ (labeled $p/q=0$ and marked by dashed lines); the
absolute conductivity is minimum (resistance maximum) close to
$90^\circ$. Less obvious are weak conductivity resonances (labeled
$p/q=\pm 1$ and marked by dotted lines) superimposed on the sloping
background associated with $p/q=0$ resonances. The most striking
aspect of all of these resonances is the fact that their positions
depend on the magnetic field strength, i.e. they do not occur at
fixed field orientations (magic angles). This result is quite
different from the DC limit, where AMRO minima are always observed
at the same field orientations for any field strength. Nevertheless,
as we will now explain, the trends observed here at high-frequency
are entirely consistent with Eqs.~\ref{eqn:6}-\ref{eqn:9}.

In order to better understand the general form of the data in
Fig.~\ref{fig:ClO4data}(a), it is important to first calibrate the
field orientation relative to the crystallographic axes. Separate
measurements to higher magnetic fields (not shown, see
\cite{SusumuPRB05}) reveal clear features associated with the
field-induced-spin-density-wave (FISDW) transition. These features
scale in field as $1/\cos\theta$ \cite{OsadaPB01} and do not shift
with frequency; $\theta$ is the angle between the applied field and
the crystallographic $c^*$ direction. In this way, one can identify
the principal symmetry directions, and these are indicated in
Fig.~\ref{fig:ClO4data}(a).

\subparagraph{Conductivity maxima (resonances).} The $c^*$-axis
conductivity ($\sigma_{zz}$) is dominated by hopping in the $c$
direction, i.e. $p=0$ and $q=1$, or $p/q=0$ (see Fig.~\ref{fig:FS}).
In low-field $c^*$-axis DC ($\omega = 0$) AMRO measurements, the
deepest resistance minima ($\sigma_{zz}$ maxima) occur when
$\omega_{01} = 0$, i.e. when $\theta=\theta_{01}$ or, equivalently,
when the field is directed along the appropriate FS warping
direction (see Fig.~\ref{fig:FS}(e) and \cite{KangSM03}), which is
$c^\prime$ ($R_{01}$) in this case [projection of $c$ onto the FS
($b^\prime c^*$-plane)]. Here, we see in fact that $\sigma_{zz}$
resonances {\em do not} occur when the field is directed along
$c^\prime$. This is due to the finite measurement frequency
(denominator in Eq.~\ref{eqn:8}). One instead sees resonances when
$\omega_{01} = \pm\omega$, i.e. either side of $c^\prime$ at angles
given by Eq.~\ref{eqn:9}. Nevertheless, in the high-field limit
$\omega_{01}^{max} >> \omega$ ($\omega/\omega_{01}^{max}\rightarrow
0$), the resonance condition reverts to the DC result. This is
apparent in Fig.~\ref{fig:ClO4data}(a) where one sees that the two
main conductivity peaks move together with increasing field such
that they should eventually superimpose at $\theta=\theta_{01}
(\equiv c^\prime)$ in the infinite field limit (see also
Fig.~\ref{fig:points} below). In addition to the $p/q=0$ resonances,
weaker $p/q=\pm 1$ resonances can be seen either side of $90^\circ$.

\subparagraph{Conductivity minima.} The $c^*$-axis conductivity
should be minimum when $|\omega - \omega_{01}|$ is maximum (see
Eq.~\ref{eqn:8}). For $\omega_{pq}^{max}>\omega>0$ there are two
such directions: (i) when $\omega_{01}=0$ ($\theta=\theta_{01}$),
corresponding to the field applied along $c^\prime$; and (ii) when
$\omega_{01}=\omega_{01}^{max}$ $[(\theta-\theta_{01})=90^\circ]$,
corresponding to the field directed $90^\circ$ away from $c^\prime$
($\cong 5^\circ$ away from $b^\prime$). The data in
Fig.~\ref{fig:ClO4data}(a) show precisely two such minima, with a
conductivity resonance in between these directions. For $\omega
>\omega_{pq}^{max}$, there is no resonance: the
conductivity nevertheless exhibits a non-resonant maximum when
$\omega_{01}=\omega_{01}^{max}$ and a minimum when $\omega_{01}=0$,
corresponding respectively to the field applied parallel and
perpendicular to $c^\prime$.

\smallskip

Further evidence that the POR {\em do not} occur at the magic angles
can be seen in Figs.~\ref{fig:ClO4data}(b) and (c), where the field
strength is varied while its orientation is fixed. Again, one
clearly observes a broad peak in $\sigma_{zz}$ in
Fig.~\ref{fig:ClO4data}(b) (labeled $p/q=0$ and marked by dashed
lines) whose position in field shifts upon varying the field
orientation. As will be seen below (Fig.~\ref{fig:points}), this
conductivity resonance corresponds to precisely the same $p/q=0$
resonance observed in Fig.~\ref{fig:ClO4data}(a). In fact, closer
inspection also reveals clear evidence for the $p/q=\pm 1$
resonances$\--$see dotted curve and expanded view in
Fig.~\ref{fig:ClO4data}(c). The very fact that one can observe these
resonances without rotating the field indicates that the POR angles
change upon varying the magnetic field strength, i.e. the POR (AC
AMRO) angles {\em are not} `magic'.

\begin{figure}[t]
\centering
\includegraphics*[width=0.55\columnwidth]{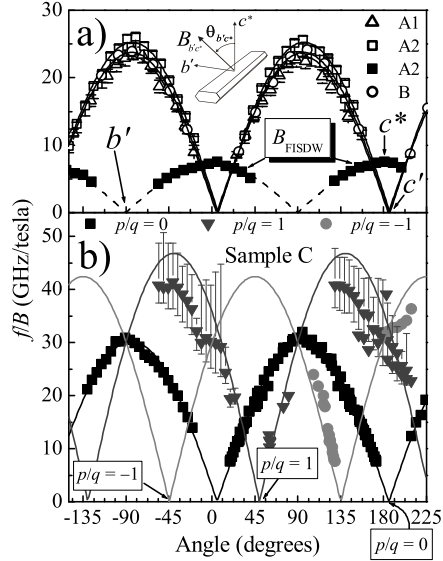}
\caption[]{Angle dependence of $f/B$ for (a) samples A and B, and
(b) sample C, for field rotation in the $b^\prime c^*$-plane; the
principal crystal axes are indicated in (a) and the different POR
have been labeled in both figures. A1 and A2 denote different cool
downs for sample A. The solid curves are fits to Eq.~\ref{eqn:7}
(see main text for detailed explanation). In (a), data are also
included for the angle dependence of the FISDW transition (solid
squares), and the inset depicts the experimental geometery.}
\label{fig:points}
\end{figure}

By combining all data obtained from the two methods, one can compile
a 2D plot of all POR data (either resonance field versus angle, or
resonance angle versus field). In fact, the relevant parameters are
the field orientation, $\theta$, and the ratio $\omega/B$ or $f/B$,
where $f$ is the microwave frequency (see Eq.~\ref{eqn:7}).
Fig.~\ref{fig:points} displays such 2D plots compiled from data
obtained at several frequencies (45 to 76~GHz) for three different
samples (A, B, and C); in every case, the field was nominally
rotated within the $b^\prime c^*$-plane. By plotting data in this
way, one can see from Eq.~\ref{eqn:7} that each series of resonances
should collapse onto a single sinusoidal arc given by $\rm A_\circ
\sin|\theta-\theta$$_{pq}|$, where the amplitude ${\rm A}_\circ =
ev_F R_{pq}/h$ gives a direct measure of the Fermi velocity
(provided the $R_{pq}$ are known). Earlier experiments on samples A
and B [Fig.~\ref{fig:points}(a)] were conducted only in the fixed
angle, swept field mode, and only $p/q=0$ resonances were observed
for these samples which were obtained from the same synthesis. As
seen in Fig.~\ref{fig:points}(a), to within the experimental error,
all of the data for samples A and B collapse onto a single curve.
Even data obtained for metallic samples cooled at different rates
(A1 and A2) lie on the same curves, in spite of the fact that the
POR linewidths differ significantly ($\tau\sim1.5-7$~ps
\cite{SusumuLT24}). Experiments on sample A were conducted in a
high-field magnet system, allowing for identification of the FISDW
transition (solid data points). The angle-dependence of $B_{\rm
FISDW}$ was used to confirm the orientation of the sample
\cite{SusumuPRB05}. Using the accepted value,
$R_{01}=13.1~\rm{\AA}$, and the value for A$_\circ~(=24\pm
1$~GHz/tesla) deduced from Fig.~\ref{fig:points}(a), a value of
$v_F=7.6\pm 0.3\times10^4$~m/s was obtained for samples A and B.

The data presented in Fig.~\ref{fig:points}(b) were obtained for a
third sample (C) taken from a separate synthesis. When fully relaxed
($\tau\sim6$~ps), this sample {\em does} show harmonic POR
corresponding to $p/q=\pm 1$ (in addition to $p/q=0$). Surprisingly,
the obtained A$_\circ = 30\pm 1$~GHz/tesla is significantly higher
for this sample, giving a value for the Fermi velocity of
$v_F=9.5\pm 0.3\times10^4$~m/s. Very preliminary studies on yet
another sample also yielded similarly high A$_\circ$ values
\cite{KovalevJAP03}. Therefore, it seems that sample quality has a
direct influence on the electronic bandwidth. This could be related
to the proximity of the SDW phase, as cleaner (more metallic)
samples exhibit higher $v_F$ values (larger bandwidth). There is
also an apparent correlation between the higher $v_F$ samples and
the observation of higher harmonic POR. Whether this is due to the
longer scattering times, or to differences in electronic structure,
is not clear. Interestingly, it has been demonstrated in a separate
study that there is a direct correlation between the scattering
time, $\tau$, and the superconducting T$_c$, suggesting
unconventional pairing in (TMTSF)$_2$ClO$_4$ \cite{SusumuLT24}.

The solid curves in Fig.~\ref{fig:points}(b) were generated using
Eq.~\ref{eqn:7} in the following way: the value of A$_\circ$, and
the $b^\prime$ and $c^\prime$ directions were determined on the
basis of fits to the $p/q=0$ POR data (solid squares); the lighter
colored curves corresponding to the $p/q=\pm1$ POR were then
simulated simply by replacing $R_{pq}$ and $\theta_{pq}$ with the
appropriate (published \cite{IYS}) values in Eq.~\ref{eqn:7}. As can
be seen, the agreement is rather good, especially at higher fields
(smaller $f/B$). The error bars increase for the weaker $p/q=\pm1$
POR at the highest $f/B$ values because the absolute uncertainty in
determining their location is relatively field independent (even
increasing slightly with decreasing field). The discrepancy at low
fields between the harmonic POR and the simulations could also
indicate a possible field dependence of the electronic bandwidth
($v_F$). However, this discrepancy is barely outside of the accuracy
of the measurement.

Each POR harmonic observed in Figs.~\ref{fig:ClO4data}
and~\ref{fig:points}(b) arises from a particular tight-binding
transfer integral $t_{pq}$. The relative amplitudes of each of the
POR harmonics are related directly to these integrals through the
coefficients, $A_{ii}^{pq}$ ($\propto t_{pq}^2$), in
Eq.~\ref{eqn:8}. The $t_{pq}$ affect the nature of the FS warping
(Eq.~\ref{eqn:5}) which, in turn, influences the tendency of the FS
to nest. A higher harmonic content to the FS warping tends to
suppress density-wave instabilities. Once again, the general trends
observed in these experiments agree with other experimental
observations, namely that the more metallic samples exhibit higher
harmonic POR (higher harmonic FS warping). Nevertheless, it is quite
clear from these measurements that interlayer hopping ($q=1$) to the
nearest-neighbor ($p=0$) sites is considerably stronger than hopping
to next-nearest-neighbor ($p=-1$) and 3$^{\rm rd}$-nearest neighbor
($p=$+1) sites$\--$see Figs.~\ref{fig:FS} and \ref{fig:ClO4data}. In
the following section, we present contrasting data for a system
showing very high harmonic POR content. Due to the significantly
higher conductivity in (TMTSF)$_2$ClO$_4$ along the $b^\prime$
direction, it has not been possible to observe $q=0$ POR.

There are some important differences between the $\sigma_{zz}$ POR
data displayed in Fig.~\ref{fig:ClO4data} and published DC AMRO data
for (TMTSF)$_2$ClO$_4$ \cite{KangSM03}. In the present work, POR are
observed in the $\omega\rightarrow0$ limit when the field is along
$c^\prime$ ($p/q=0$) and roughly $\pm45^\circ$ away from $c^\prime$
($p/q=\mp1$). DC measurements reveal very clear evidence for higher
harmonic AMRO such as $p/q=\pm2$ and $\pm3$. However, these are only
seen in regions of parameter space that were inaccessible with the
instrumentation used for the microwave measurements. Comparisons of
interlayer ($q=1$) POR and AMRO data obtained over comparable
temperature, field and field orientation ranges reveal similar
behaviors in terms of the relative strengths of the $p/q=0$ and $\pm
1$ resonances. However, a surprising aspect of the DC AMRO
measurements is a strong, sharp $q=0$ ($p/q=\infty$) resonance when
the field is parallel to $b^\prime$ \cite{KangSM03}, though this dip
is less obvious in separate earlier studies \cite{Chash97}. There is
no evidence for such a resonance in the microwave measurements. The
reason for this difference is not entirely clear. According to
Eq.~\ref{eqn:8}, $\sigma_{zz}$ should not exhibit any $q=0$
resonances, since $A_{zz}^{pq}=0$ for $q=0$. One possibility may be
that the DC $\sigma_{zz}$ measurements are contaminated with
$\sigma_{yy}$, which should display a dominant $p/q=\infty$
resonance. However, this may suggest additional (perhaps
non-classical) effects in the DC conductivity which do not influence
the microwave conductivity \cite{CSCC}.

\subsection{$\alpha$-(BEDT-TTF)$_2$KHg(SCN)$_4$}
\label{sec:6}

Like the Bechgaard salts, the Q2D
$\alpha$-(BEDT-TTF)$_2$MHg(SCN)$_4$ (M = NH$_4$, K, Tl, Rb) family
of organic charge-transfer salts have a rich history in terms of
studies of their Fermiology
\cite{IYS,Wosnitza,JSReview,KartsReview,BlundPRB96,JSPRL92,HillSM93,HillSM95,Poliskii,HillPRB97,BlundPRB97,ArdPRL98,ArdPRB99,KovalevPRB02}.
Indeed, the first CR studies were performed on the M~=~K member of
this family \cite{JSPRL92}, which is also the focus of the present
section. For this compound, the least conducting ($z$-) direction is
parallel to the crystallographic $b^*$-axis \cite{IYS,Mori90}.
Meanwhile, the conductivity within the $ac$-plane ($\perp b^*$) is
rather more isotropic than the conductivity within the $ab$-plane
for (TMTSF)$_2$ClO$_4$. Consequently, the room temperature FS of
$\alpha$-(BEDT-TTF)$_2$KHg(SCN)$_4$ has a more two-dimensional
character \cite{Mori90,Campos96}, comprising both open (Q1D) and
closed (Q2D) pockets (see Fig.~1 in Ref. \cite{KovalevPRB02}). At
low temperatures, matters are complicated by the fact that
$\alpha$-(BEDT-TTF)$_2$KHg(SCN)$_4$ undergoes a charge-density-wave
(CDW) transition at 8~K \cite{IYS,NeilPRL99}, which leads to a
reconstruction of the room temperature FS
\cite{Karts93,Wosnitza,JSReview,KartsReview,NeilJPCM99}. Unlike the
anion ordering in (TMTSF)$_2$ClO$_4$, the low temperature state in
$\alpha$-(BEDT-TTF)$_2$KHg(SCN)$_4$ is not sensitive to the cooling
rate through this CDW transition. The precise nature of the
reconstructed FS remains controversial. However, it is believed that
the open sections nest, and that the closed pockets reconstruct in
such a way as to give rise to new open FS sections together with
smaller closed pockets \cite{Karts93,NeilJPCM99}; one of the
original proposals for the reconstructed FS is shown in Fig.~1 of
Ref.~\cite{KovalevPRB02}. This model serves to illustrate most
aspects of the POR data presented in this section.

Due to the uncertainty associated with the low temperature FS,
several early magnetooptical studies of
$\alpha$-(BEDT-TTF)$_2$KHg(SCN)$_4$ incorrectly attributed resonant
absorptions to the conventional CR phenomenon
\cite{JSPRL92,HillSM93}. The first indications that they were in
fact due to open orbit POR came from angle-dependent cavity
perturbation measurements by Ardavan et al. \cite{ArdPRL98}. In
these investigations, the applied magnetic field was rotated in two
different planes perpendicular to the highly conducting $ac$-plane,
which is easily identified from the plate-like shape of a typical
single crystal. In the case of Q2D CR, the cyclotron frequency
depends only on the magnitude of the field component perpendicular
to the conducting layers and should, therefore, be insensitive to
the particular plane of rotation \cite{HillPRB97}. The main clue
that the resonances observed in $\alpha$-(BEDT-TTF)$_2$KHg(SCN)$_4$
were due to Q1D POR came from the fact that the angle dependence
varied strongly with the plane of rotation \cite{BlundPRB97}. In
this study, Ardavan et al., were able to fit data corresponding to
two FS warping components to sinusoidal arcs of the form given by
Eq.~\ref{eqn:7} \cite{ArdPRL98}.

\begin{figure}[t]
\centering
\includegraphics*[width=0.75\columnwidth]{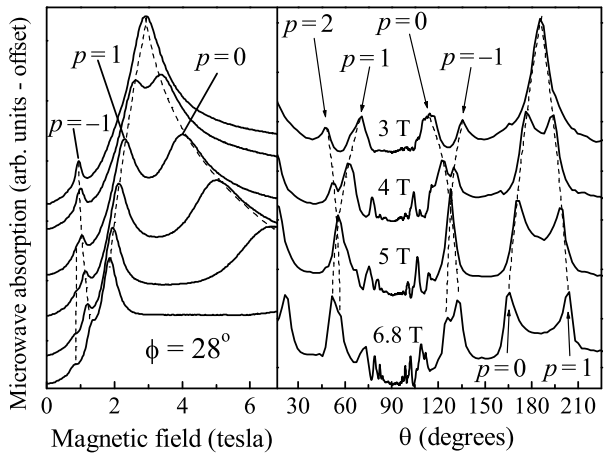}
\caption[]{Microwave absorption data ($\propto\sigma_{zz}$) for
$\alpha$-(BEDT-TTF)$_2$KHg(SCN)$_4$ obtained in (a) field- and (b)
angle-swept modes (from \cite{KovalevPRB02}). All data were obtained
at 2.2~K and 53.9~GHz for field rotation in a plane inclined at
$\phi=28^\circ$ with respect to the Q1D FS, and the traces have been
offset for clarity. The POR, corresponding to peaks in absorption,
have been labeled according to the ratio of $p/q$.}
\label{fig:KHgdata}
\end{figure}

Improvements in spectrometer design (enhanced sensitivity and
mechanical stability \cite{MolaRSI,SusumuRSI}) have since enabled
cavity perturbation measurements with improved signal-to-noise
characteristics. Fig.~\ref{fig:KHgdata} displays microwave loss data
obtained by Kovalev et al. \cite{KovalevPRB02} at 2.2~K and
53.9~GHz, both in field-swept (a) and angle-swept (b) mode. Similar
to (TMTSF)$_2$ClO$_4$, the sample was positioned within the cavity
so as to excite interlayer ($b^*$-axis) currents throughout the bulk
of the sample \cite{HillPRB00}. Thus, the vertical scale in
Fig.~\ref{fig:KHgdata} is proportional to $\sigma_{zz}$. Unlike
(TMTSF)$_2$ClO$_4$, however, the reduced in-plane anisotropy,
together with the 8~K FS reconstruction, make it impossible to
visually identify the orientation of the open FS associated with the
low-temperature state. Consequently, one does not initially know the
relative angle between the field rotation plane and the Q1D FS
($xz$-plane). It is therefore necessary to extend the theory
developed in section~\ref{sec:3} for arbitrary rotations
\cite{KovalevPRB02}, i.e. not just rotations within the plane of the
FS. Before doing so, however, we present a detailed discussion of
the data in Fig.~\ref{fig:KHgdata}.

The first point to note upon comparison with Fig.~\ref{fig:ClO4data}
is the significant increase in the harmonic content of the POR for
$\alpha$-(BEDT-TTF)$_2$KHg(SCN)$_4$. This is particularly apparent
from the angle-swept data. Indeed, careful inspection of
Fig.~\ref{fig:KHgdata}(b) reveals over 17 clear resonances within a
$180^\circ$ angle range, along with an apparent continuum of peaks
as the field orientation approaches $90^\circ$ ($\perp$ to $b^*$);
several of these peaks have been labeled. This behavior is
dramatically different from that observed for (TMTSF)$_2$ClO$_4$,
and is indicative of a high harmonic content to the FS warping
\cite{BlundPRB96}. However, it is not unexpected, as DC AMRO data
are equally rich \cite{JSReview}. As will be seen below, this
behavior is related to the intrinsic quasi-two-dimensionality of
this BEDT-TTF compound, along with the 8~K CDW transition and the
resulting FS reconstruction \cite{Karts93,NeilJPCM99}.

There are certain similarities between Fig.~\ref{fig:ClO4data}(a)
and Fig.~\ref{fig:KHgdata}(b), but also some key differences. First
of all, both figures exhibit a broad conductivity minimum when the
applied field is perpendicular to the least conducting direction
($90^\circ$). Furthermore, the POR patterns for both materials
exhibit symmetry about $\theta=90^\circ$. In contrast, the behavior
around (or close to) $\theta=0^\circ$ and $180^\circ$ is quite
different for the two materials. As can be seen from
Fig.~\ref{fig:ClO4data}(a), the two $p/q=0$ resonances either side
of $\theta=0$ diverge as $B\rightarrow0$ ($\omega>>\omega_{01}$
limit), whereas they appear to merge together in the high field
($\omega/\omega_{01}\rightarrow0$) limit, suggesting that the
warping direction in (TMTSF)$_2$ClO$_4$ is along $c^\prime$
($5^\circ$ away from $c^*$); in other words, the relevant warping
direction corresponds to nearest-neighbor interlayer hopping. By
contrast, the two peaks seen in Fig.~\ref{fig:KHgdata}(b) merge at
3T. In fact, these two peaks do not even correspond to the same POR
harmonic. If one looks carefully at the labeling in
Fig.~\ref{fig:KHgdata}(b) ({\em vide infra}), it is apparent that
the $p/q=0$ peaks will merge somewhere around $\theta=150^\circ$.
This implies that the relevant $R_{01}$ tight-binding hopping
direction possesses a significant in-plane component. We note that
none of the observed POR have a $\omega/\omega_{pq}\rightarrow0$
intercept anywhere close to $\theta=0$, implying that all relevant
interlayer hopping matrix elements possess significant in-plane
components. Even though $\alpha$-(BEDT-TTF)$_2$KHg(SCN)$_4$
possesses a low-symmetry triclinic ($P\bar{1}$) structure, this
degree of obliqueness of the $R_{pq}$ vectors $cannot$ be explained
from high temperature (104~K) crystallographic data \cite{Mori90}.
We shall discuss this point further below, along with the
observation that the POR imply a very weak decay of the $t_{pq}$
with increasing $|p|$.

As discussed above, before we can fit the POR peak positions to the
model developed in section~\ref{sec:3}, we must modify
Eq.~\ref{eqn:7} slightly for an arbitrary plane of rotation of the
applied field. We also adapt this equation and Eq.~\ref{eqn:6} so
that they are appropriate for the
$\alpha$-(BEDT-TTF)$_2$KHg(SCN)$_4$ crystal structure [see
Fig.~\ref{fig:FS}(e)], for which $b^*$ corresponds to the least
conducting $z$-direction. For field rotation in a plane inclined at
an angle $\phi$ with respect to the FS ($yz$-plane), the DC AMRO
condition is given by,

\begin{equation}\label{eqn:10}
\tan \theta _{pq}  = \frac{{1}}{{\cos\phi}}\left[\frac{{pl}}{{qb^*
}} + \frac{l^\prime}{b^*}\right],
\end{equation}

\noindent{where, as usual}, the angles $\theta_{pq}$ are measured
relative to the $z$-axis (see also Fig.~\ref{fig:FS}(e) for a
definition of the parameters $l$ and $l^\prime$). The modified
resonance condition is then given by the following expression:

\begin{equation}\label{eqn:11}
\frac{\omega_{pq}}{B\cos\theta}  = \frac{{v_F eb^*}}{\hbar}
\cos\phi|\tan\theta_{pq}-\tan\theta|.
\end{equation}

\noindent{From} Eq.~\ref{eqn:11} it can be seen that plots of
$f/B\cos\theta$ versus $\tan\theta\cos\phi$ should produce straight
lines with slope $\pm e b^* v_F/h$, with offsets given by
$\theta_{pq}$. Such a plot is shown in Fig.~\ref{fig:KHgscaling} for
data obtained for two planes of rotation ($\phi=28^\circ$ and
$\phi=66^\circ$). As can clearly be seen, the data scale well,
particularly in the lower central portion of the plot corresponding
to field orientations well away from $\theta=90^\circ$. These
results therefore confirm the Q1D nature of the POR, as previously
reported by Ardavan et al. \cite{ArdPRL98}. The solid lines
represent the best fit to the data with $v_F$, $l$ and $l^\prime$ as
the only free parameters. The ratios $l/b^*=1.2$ and
$l^\prime/b^*=0.5$ are in excellent agreement with DC AMRO
measurements \cite{KovalevSSC94,Iye94,Jason95}. The obtained value
of $v_F=6.5\times 10^4$~m/s, meanwhile, cannot be deduced from DC
AMRO measurements. The deviation between the data and the fit in the
peripheral regions of Fig.~\ref{fig:KHgscaling} may have several
explanations. First of all, errors in the calibration of the field
orientation will be amplified for small values of $\cos\theta$ and
large values of $\tan\theta$. It is also likely that the above
theory breaks down when the field is oriented close to the direction
perpendicular to the plane of the Q1D FS (small $\cos\theta$,
$\cos\phi$ and large $\tan\theta$) due to the Q2D nature of
$\alpha$-(BEDT-TTF)$_2$KHg(SCN)$_4$, i.e. one can no longer assume
that $v_x$ is a constant of the motion, resulting in
non-separability of all three velocity components. One may estimate
an effective mass associated with the high temperature Q2D FS from
the obtained value of $v_F$. However, such a procedure involves
making assumptions about the energy dispersion. Nevertheless, the
obtained effective mass ($m^*=1.6-2.4m_e$, depending on the
assumption made about the energy dispersion) is in reasonable
agreement with SdH and dHvA measurements \cite{JSReview} (see
\cite{KovalevPRB02} for a more in depth discussion).

\begin{figure}[t]
\centering
\includegraphics*[width=0.6\columnwidth]{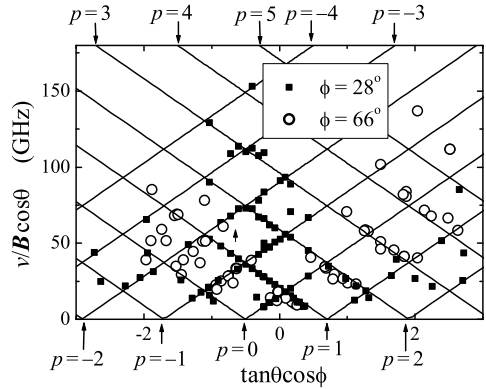}
\caption[]{A compilation of all data obtained for
$\alpha$-(BEDT-TTF)$_2$KHg(SCN)$_4$ scaled according to
Eq.~\ref{eqn:11} (from \cite{KovalevPRB02}). Several branches have
been labeled with the appropriate index $p$.} \label{fig:KHgscaling}
\end{figure}

We end this section with a discussion of the very high harmonic
content of the POR observed for $\alpha$-(BEDT-TTF)$_2$KHg(SCN)$_4$.
If one takes the view that each resonance corresponds to a given
tight-binding transfer integral, then the present study (as well as
DC AMRO measurements \cite{JSReview,BlundPRB96}) would imply a very
weak decay of the $t_{pq}$ with increasing $|p|$ (for $q=1$). Such a
result would not justify the use of a tight-binding approximation,
as it would imply significant orbital overlap to next-nearest and
next-next-nearest, etc. sites in the lattice [see
Fig.~\ref{fig:FS}(e)]. However, as pointed out by Blundell et al.
\cite{BlundPRB96}, such a view is inappropriate for the
low-temperature FS in $\alpha$-(BEDT-TTF)$_2$KHg(SCN)$_4$ due to the
8~K reconstruction that results from the CDW superstructure
\cite{Karts93,NeilJPCM99}. The more appropriate view to take is that
each POR harmonic corresponds to a fourier component of the warping
of the Q1D FS. Interestingly, inspection of Fig.~\ref{fig:KHgdata}
reveals similar amplitudes for the $p=0$ and 1 POR, which have
similar $|R_{pq}|$ (they differ in amplitude by less than $10\%$);
the same is true for the $p=2$ and $-1$ POR, for which $|R_{21}|$
and $|R_{-11}|$ differ by less than $7\%$. Thus, it would appear as
though the POR intensities scale as some power law of the associated
$R_{pq}$ ($\propto b^*/|\cos\theta_{pq}|$). Meanwhile, the
corresponding fourier amplitudes, $t_{pq}$, scale as the square root
of the POR intensities, i.e. $A_{zz}^{pq}\propto t_{pq}^2$, see
Eq.~\ref{eqn:8}. We note that for the most extreme anharmonicity,
the top hat (or square wave) function, the $t_{pq}$ should scale as
$R_{pq}^{-1}$ and the POR intensities as $R_{pq}^{-2}$: $R_{01}$ and
$R_{21}$ differ by roughly a factor of 2 [$(R_{01}/R_{21})^2=4$],
while the corresponding $p/q=0$ and $p/q=2$ POR differ in intensity
by about a factor of 6. Thus, the POR intensities imply an extremely
high harmonic content to the FS warping, which can only be explained
in terms of a reconstruction. It is also clear that the
$\vec{R_{pq}}$ bear no simple relationship with the principal
lattice vectors. Again, this is because the corrugations on the
reconstructed FS are related to the CDW nesting vector, not the
underlying lattice structure. Therefore, POR measurements are
entirely consistent with proposed models for the low-temperature
Fermi surface of $\alpha$-(BEDT-TTF)$_2$KHg(SCN)$_4$
\cite{Karts93,NeilJPCM99}.

\section{Open-orbit POR in a Q2D system} \label{sec:9}

We conclude this experimental survey by describing a new open-orbit
POR effect which was recently reported by Kovalev et al. in the Q2D
organic conductor $\kappa$-(BEDT-TTF)$_2$I$_3$ \cite{KovalevPRL03}.
The new effect is observed when the magnetic field is applied
parallel to the highly conducting layers$\--$the $bc$-plane for this
particular compound. As before, the Lorentz force induces periodic
quasiparticle trajectories on the warped FS such that
$\vec{\dot{k}}$ is perpendicular to the applied field. However, in
this case, the FS is a warped cylinder, as depicted in
Fig.~\ref{fig:I3}(a). Nonetheless, because the FS is warped (finite
dispersion along $a$), the Lorentz force leads to periodic
modulations of the interlayer quasiparticle velocities, $v_z$, which
in turn give rise to resonant contributions to $\sigma_{zz}$ (see
Eq.~\ref{eqn:3}). As can be seen from Fig.~\ref{fig:I3}(a), a range
of different trajectories are induced: open-orbits, self-crossing
orbits and closed orbits
\cite{KartsReview,Hanasaki98,PeschPRB99,JSPRL02}. Furthermore, since
$\vec{\dot{k}}$ is proportional to $(\vec{v_F}\times\vec{B})$, it is
apparent that the periodicities associated with each orbit will vary
significantly over the FS, i.e. the POR frequencies associated with
different states on the FS are not discrete, as was the case in the
previous examples. The DC $\sigma_{zz}$ (AMRO) is believed to be
dominated by the self-crossing orbits, though this has been the
subject of some debate \cite{Hanasaki98,PeschPRB99}.

\begin{figure}[t]
\centering
\includegraphics*[width=0.95\columnwidth]{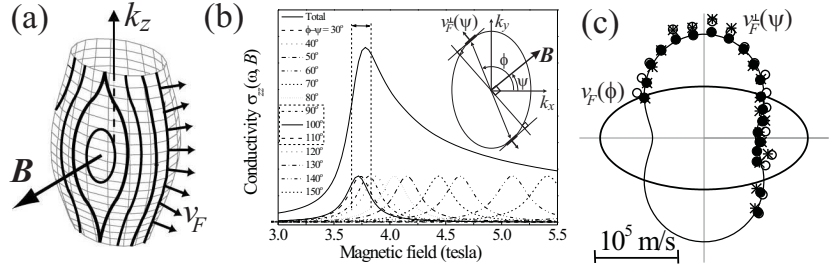}
\caption[]{(a) An illustration of the quasiparticle trajectories on
a warped Q2D FS cylinder for a field oriented perpendicular to the
cylinder axis. The resulting trajectories lead to $v_z$ oscillations
and to a resonance in $\sigma_{zz}$ (see main text). The solid curve
in (b) illustrates the conductivity resonance resulting from the
electron trajectories in (a). Different parts of the FS contribute
to different parts of the resonance (dashed curves). However, the
density of resonances is highest for FS patches which are parallel
to the applied field (see inset). Consequently, these FS regions
dominate the POR. In (c), experimental data are plotted
corresponding to $v_F^\perp$ obtained by various different methods
(see \cite{KovalevPRL03} for explanation); the dashed curve is a fit
to the data and the solid curve represents the corresponding
$v_F(\phi)$.} \label{fig:I3}
\end{figure}

The situation at microwave frequencies is considerably simpler
(provided $\omega\tau>1$). As illustrated in Fig.~\ref{fig:I3}(b),
$\sigma_{zz}(\omega)$ exhibits a peak (quasi-resonance) which is
dominated by contributions from quasiparticle states corresponding
to vertical (symmetry equivalent) strips of the FS which are tangent
to the applied field, as depicted in the inset to
Fig.~\ref{fig:I3}(b). The reason for the dominance of these
trajectories is two-fold. First of all, every periodic trajectory
has an associated frequency $\omega_c(\vec{k})$. However, these
frequencies vary from zero (smallest closed orbits) up to a maximum
cut-off frequency, $\omega_c^{max}$, given by the maximum value of
$|(\vec{v_F}\times\vec{B})|$. This cut-off gives rise to a
singularity in $\sigma_{zz}(\omega)$. Note that $\omega_c^{max}$
corresponds to states with $\vec{v_F}\perp\vec{B}$, i.e. states on
FS sections which are tangent to $\vec{B}$. Second, the density of
momentum states diverges at the cut-off frequency, i.e. the number
of states per unit frequency range diverges for the tangential
patches. The resultant conductivity is obtained by integration over
all states on the FS (or all frequencies from 0 to
$\omega_c^{max}$). As illustrated in Fig.~\ref{fig:I3}(b), such an
integration gives rise to a conductivity resonance. The actual value
of $\omega_c^{max}$ ($=eBav_F^\perp/\hbar$, where $a$ is the
interlayer spacing and $v_F^\perp$ is the Fermi velocity associated
with the tangential patches) corresponds to the point on the curve
having the maximum slope, i.e. slightly to the low-field side of the
peak in $\sigma_{zz}(B)$. Measurement of $\omega_c^{max}$ as a
function of the field orientation $\psi$ within the $xy$-plane
yields a plot of $v_F^\perp(\psi)$. The procedure for mapping
$v_F^\perp(\phi)$ is then identical to that of reconstructing the FS
of a Q2D conductor from the measured periods of Yamaji oscillations
\cite{IYS,Yamaji} (see \cite{KovalevPRL03} for more in-depth
discussion).

Actual experimental data corresponding to $v_F^\perp(\psi)$, along
with the deduced $v_F^\perp(\phi)$, are displayed in
Fig.~\ref{fig:I3}(c). As with all of the previous POR
investigations, a cavity perturbation technique was employed in such
a way as to excite only interlayer currents
\cite{HillPRB00,MolaRSI}. The interlayer ($a$) direction is easily
identified due to the platelet shape of a typical
$\kappa$-(BEDT-TTF)$_2$I$_3$ single-crystal. Experiments were
performed at $4.5$~K (above the superconducting transition
temperature of 3.5~K) and at a frequency of 53.9~GHz, corresponding
to the TE011 mode of the employed cavity. Use of both the
transmitted phase and amplitude enabled precise determination of
$\omega_c^{max}(\phi)$ (see \cite{KovalevPRL03} for representative
spectra) and the scattering time $\tau$ (=~5~ps). The FS of
$\kappa$-(BEDT-TTF)$_2$I$_3$ consists of a network of overlapping
weakly warped cylinders. The underlying lattice periodicity results
in a removal of the degeneracies at the intersections of these
cylinders, giving rise to the coexistence of smaller closed surfaces
and open sheets. The deduced anisotropy in $v_F^\perp(\phi)$ is in
good agreement with the known anisotropy associated with small Q2D
FS for $\kappa$-(BEDT-TTF)$_2$I$_3$ \cite{Kobayashi95}, i.e.
$v_F^x=1.3\times10^5$~m/s and $v_F^y=0.6\times10^5$~m/s.
Furthermore, one can estimate the effective mass associated with
this Q2D pocket using the relation $m^*=\hbar(S_k/S_v)^{1/2}$, where
$S_k$ is the cross sectional area of the FS in $k$-space and $S_v$
the area enclosed by $v_F^\perp(\phi)$ in 2D velocity space. This
procedure gives $m^*=1.7m_e$ \cite{KovalevPRL03}, while the
experimental value deduced from SdH and dHvA measurements is $\sim
1.9m_e$ \cite{Balthes}.

\subsection{POR in Q2D nodal superconductors} \label{sec:10}

In the case of the high-T$_{\rm c}$ superconductors and other
candidate Q2D $d$-wave superconductors (including several organic
conductors \cite{RossScience}), it is generally accepted that normal
quasiparticles coexist with the superfluid along vertical line-nodes
on the original approximately cylindrical high-temperature FS
\cite{TsueiRMP} (see Fig.~\ref{fig:nodes}). These quasiparticles
will dominate the low temperature low-energy electrodynamics,
including the microwave spectral range (all other single-particle
excitations are gapped) \cite{Hosseini,CorsonPRL00,TurnerPRL03}. A
magnetic field applied parallel to the $xy$-plane [$B(\psi)$]
preserves in-plane momentum. Consequently, such a field will tend to
drive quasiparticles along the vertical line nodes, thus preserving
the open orbit POR effect described above. What is more, the nodal
quasiparticles will tend to be even longer lived than in the normal
state due to the reduced phase space for scattering \cite{Hosseini}.
Therefore, this dominance of the nodal regions of the FS suggests
that it may be possible to directly probe quasiparticles in nodal
superconductors via open-orbit POR.

\begin{figure}[t]
\centering
\includegraphics*[width=0.85\columnwidth]{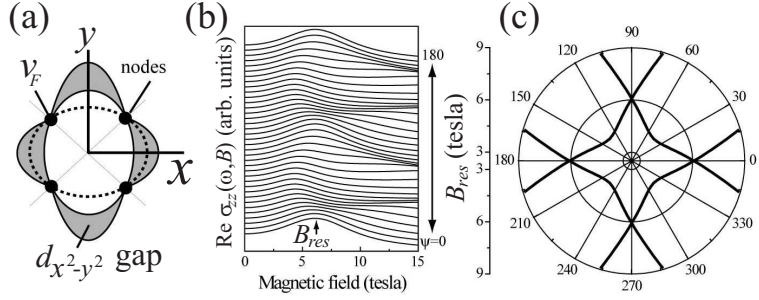}
\caption[]{(a) Fermi surface and gap for a $d_{x^2-y^2}$
superconductor, along with (b) the corresponding calculation of the
interlayer AC conductivity, $\sigma_{zz}(\psi,B)$, and (c) the angle
dependence of the resulting resonance fields, $B_{res}$ (from
\cite{SusumuJAP05}). Because of the assumption to count only the
contribution from normal quasiparticles at the nodes, a discrete
$v_F(\phi)$ was considered (see \cite{SusumuJAP05} for further
details).} \label{fig:nodes}
\end{figure}

Fig.~\ref{fig:nodes}(a) illustrates the situation discussed above
for a system with a warped elliptical FS (only the cross-section is
shown), such as $\kappa$-(BEDT-TTF)$_2$X (X=I$_3$, Cu(NCS)$_2$, etc.
\cite{IYS}), with a $d_{x^2-y^2}$ superconducting gap (shaded area);
the dashed curve represents the corresponding angle-dependent Fermi
velocity, $v_F$. Within the superconducting state, the FS is gapped
everywhere, apart from at the locations of the four line nodes.
Fig.~\ref{fig:nodes}(b) shows numerical simulations of the field
dependent interlayer ($z$-axis) conductivity, $\sigma_{zz}(\psi,B)$,
for different field orientations, due to the quasiparticles existing
at the line nodes; refer to \cite{SusumuJAP05} for further details
of these calculations. The peaks in conductivity correspond to the
open-orbit POR described in the previous section. However, the
angle-dependence is now dominated by the line nodes, rather than the
extremal regions of the FS. Thus, the conductivity exhibits two
resonances corresponding to pairs of line-nodes on opposite sides of
the FS. The resonance field $B_{res}(\psi)$ depends simply on the
angle between the applied field and the lines joining these
line-node pairs$\--$hence the two resonances. As shown in
Fig.~\ref{fig:nodes}(c), $B_{res}(\psi)$ exhibits a four-fold
symmetry characteristic of the $d_{x^2-y^2}$ gap, as opposed to the
two fold symmetry characteristic of the original un-gapped
elliptical FS [Fig.~\ref{fig:I3}(c)].

In principle, one could apply this POR technique to any nodal
superconductors which satisfy $\omega\tau>1$. Using this method, it
may be possible to measure both $v_F$ (effective mass) and $\tau$
associated with nodal quasiparticles. Moreover, it may also be able
to confirm the symmetry of the superconducting gap from the
angle-dependence of the POR. The organic superconductors may be
particularly attractive for such investigations due to their extreme
purity. However, recent experiments suggest that it may also be
possible to observe this effect in very pure high-T$_{\rm c}$
compounds such as Y$_2$Ba$_2$Cu$_3$O$_{6+x}$ \cite{TurnerPRL03} and
Tl$_2$Ba$_2$CuO$_{6+\delta}$ \cite{Hussey}.

\section{Discussion and comparisons with other experiments} \label{sec:11}

\begin{table}[!pb]
\centering \caption[]{Physical parameters deduced from these
investigations and from the literature for: {\bf (1)}
(TMTSF)$_2$ClO$_4$; {\bf (2)} $\alpha$-(BEDT-TTF)$_2$KHg(SCN)$_4$;
and {\bf (3)} $\kappa$-(BEDT-TTF)$_2$I$_3$. The transport scattering
time ($\tau$) and Fermi velocity ($v_F$) were deduced from POR
measurements. The effective masses ($m^*$) for {\bf (2)} and {\bf
(3)} were deduced from the obtained values of $v_F$, as described in
refs \cite{KovalevPRB02} and \cite{SusumuPRB05}, respectively; $m^*$
for {\bf (1)} was deduced on the basis of a tight binding model, as
described in \cite{KovalevPRB02}. The quantity $ne^2\tau/m^*$
represents the DC conductivity deduced on the basis of the $\tau$
and $m^*$ values obtained from the POR measurements, using
literature values for the carrier concentration $n$
\cite{IYS,Mori90,Tamura91}; we assume that a carrier density
corresponding to only $20\%$ of the Brillouin zone contributes to
the conductivity in the low-temperature state of {\bf (2)}
\cite{JSReview}. $\tau_h$ is the interlayer hopping time deduced
from: $a-$the interlayer bandwidth \cite{IYS}; $b-$Ref.
\cite{KovalevPRB02}, $c-$Ref. \cite{WosnitzaPRB02}. $\tau_\varphi$
is the quantum lifetime deduced from magneto oscillation data:
$d-$Ref. \cite{JSReview}, $e-$Ref. \cite{WosnitzaPRB02}. The optical
scattering time ($\tau^{opt}$) and DC conductivity
($\sigma_\circ^{opt}$) were taken from Refs.
\cite{Schwartz98,DresselPRB05} for {\bf (1)} at 10~K, from Ref.
\cite{DresselPRL03} for {\bf (2)} at 6~K, and from Ref.
\cite{Tamura91} for {\bf (3)} at 15~K.} \label{tab:1}
\renewcommand{\arraystretch}{1.2}
\setlength\tabcolsep{4.3pt}
\begin{tabular}{@{}lcccccccc@{}}
\hline\noalign{\smallskip}
& $\tau$ & $v_F$ & $m^*$ & $ne^2\tau/m^*$ & $\tau_h$ & $\tau_\varphi$ & $\tau^{opt}$ & $\sigma_\circ^{opt}$ \\
& (ps) & (/10$^4$ ms$^{-1}$) & ($m_e$) & (S cm$^{-1}$) & (ps) & (ps) & (ps) & (S cm$^{-1}$) \\
\hline\noalign{\smallskip}
{\bf 1.} & 1-7$^*$ & 7.6-9.5$^*$ & 1.45-1.75$^*$ & $\sim 10^6$ & $0.7^a$ & - & $>10$ & $10^3$-$10^4$ \\
{\bf 2.} & 15 & 6.5 & 1.6-2.4 & $3\times10^5$ & $0.75^b$ & 0.8-$2.5^d$ & 1 & $\sim 10^3$ \\
{\bf 3.} & 5 & 10 & 1.7-2.5 & $4 \times 10^5$ & $1^c$ & $5^e$ & $\sim 0.1$ & $\sim 1500$ \\
\hline
\end{tabular}
\begin{flushleft}
$^*$ cooling rate dependent \end{flushleft}
\end{table}

The first four columns in Table \ref{tab:1} summarize optical
constants deduced from the POR studies outlined in the previous
sections, for: {\bf (1)} (TMTSF)$_2$ClO$_4$; {\bf (2)}
$\alpha$-(BEDT-TTF)$_2$KHg(SCN)$_4$; and {\bf (3)}
$\kappa$-(BEDT-TTF)$_2$I$_3$. The final four columns list various
physical parameters obtained from the literature; unless otherwise
indicated, these parameters represent low-temperature limiting
values. The first point to note is that all three materials appear
to display coherent 3D band transport at low temperatures. This can
be seen by comparing the scattering times (either the transport
time, $\tau$, or the quantum lifetime, $\tau_\varphi$) with the
interlayer hopping times, $\tau_h$, i.e. for all materials,
$\tau_h<\{\tau,\tau_\varphi\}$. However, for all three materials,
the difference is not so great (roughly an order of magnitude).
Thus, one may expect a crossover to an incoherent regime at fairly
low temperatures. For ({\bf 3}), the transport lifetime deduced from
the POR measurements is in excellent agreement with the quantum
lifetime deduced from dHvA studies \cite{WosnitzaPRB02}, in spite of
the fact that the samples were grown completely independently by
different groups. For ({\bf 2}), meanwhile, there is a significant
discrepancy between the transport and quantum lifetimes. Part of the
reason for this discrepancy could be related to the complexities of
the low temperature state of $\alpha$-(BEDT-TTF)$_2$KHg(SCN)$_4$
\cite{JSReview}. In particular, the reconstructed FS consists of
multiply connected Q1D and Q2D sections, giving rise to an extremely
rich pattern of quantum oscillations involving magnetic breakdown.
It is also notable that the POR data presented in section
\ref{sec:6} clearly originate from quasiparticles belonging to an
open FS, whereas quasiparticles responsible for quantum oscillatory
phenomena necessarily involve closed orbits. Thus, a direct
comparison between $\tau$ and $\tau_\varphi$ is probably
inappropriate for ({\bf 2}). Nevertheless, a significant (order of
magnitude) difference between these quantities has previously been
noted for another member (M = NH$_4$) of the $\alpha$-phase salts,
which {\em does not} undergo a low temperature FS reconstruction
\cite{HillPRB97}. Such a difference can be explained in terms of an
inhomogeneous broadening of Landau levels, which would show up in
$\tau_\varphi$, but not necessarily in the transport lifetime
\cite{Shoen}. Due to the absence of closed pockets, a comparison
between $\tau$ and $\tau_\varphi$ is not possible for ({\bf 1}).

As discussed in section \ref{sec:2}, $\omega$ and $\omega_c$ play
essentially the same role in Eq.~\ref{eqn:4}. It is, therefore,
interesting to compare optical constants obtained from these
measurements with published values deduced from more conventional
optical methods. There is one caveat, however: most optical
reflectivity measurements are limited to frequencies well above the
scattering rates ($\tau^{-1}$) deduced from these POR studies. For
example, optical studies of ({\bf 2}) are limited to frequencies
above 50~cm$^{-1}$ (1.5~THz) \cite{DresselPRL03}, and those of ({\bf
3}) to above 500~cm$^{-1}$ (15~THz) \cite{Tamura91}. Consequently,
most of the low-temperature Drude spectral weight does not even
contribute to the measured reflectivity in these investigations. The
only material for which low-frequency cavity perturbation data do
exist (down to $\sim5$~cm$^{-1}$) is (TMTSF)$_2$ClO$_4$
\cite{Schwartz98,DresselPRB05,Ng,CaoJPI96}. Nevertheless, one still
has to rely on Kramers-Kronig and Hagen-Rubens extrapolation for
$\omega\rightarrow0$ in order to recover the Drude spectral weight
\cite{DressGrunBook}. Thus, such techniques ultimately rely on
accurate measurements of the DC conductivity. It is perhaps not
surprising, therefore, that the optical scattering rates,
$\tau^{opt}$, deduced for ({\bf 2}) and ({\bf 3}) are roughly half
the minimum frequencies employed in the measurement, i.e.
$\sim$30~cm$^{-1}$ and $\sim$300~cm$^{-1}$, respectively. In both
cases, the deduced values of $\tau^{opt}$ are nowhere near the
scattering rates deduced from POR studies. In addition, there is at
least a two orders of magnitude difference between the DC
conductivities found in optical studies and those deduced from POR
measurements which assume a $100\%$ contribution of free carriers to
the Drude peak.

For (TMTSF)$_2$ClO$_4$, the situation is only slightly better. The
values for $\tau^{opt}$ and $\sigma_\circ^{opt}$ listed in
Table~\ref{tab:1} were taken from the more recent literature
\cite{Schwartz98,DresselPRB05}. While the value of $\tau^{opt}$ is
in reasonable agreement with the POR data (because of the lower
frequencies employed), the value of $\sigma_\circ^{opt}$ clearly is
not; again, there is a factor of 100 difference between the upper
range for $\sigma_\circ^{opt}$ and the POR estimate. However, there
exist earlier reports of considerably higher conductivities,
including the original study of Bechgaard et al. \cite{Bech81},
where low temperature values in the $10^5-10^6$~S~cm$^{-1}$ range
were found. This agrees nicely with the POR value quoted in Table
\ref{tab:1}. It is notable that, when the larger conductivity is
used to extrapolate optical data, an anomalously narrow low-energy
Drude peak is obtained (width = 0.034~cm$^{-1}$, or 1~GHz)
\cite{CaoJPI96}. This highlights the problems associated with
characterization of the low energy Drude spectral weight from
conventional broad band optical techniques which are ordinarily
limited to frequencies above about 10~cm$^{-1}$. However, several
recent studies have suggested a general trend among many of the
layered organic conductors: namely, that only a small fraction (as
little as $1\%$) of the overall spectral weight is to be found in
the low-energy free carrier (Drude) response. If this is indeed the
case, then the DC conductivities deduced from POR measurements
should be revised downwards by a factor of 100, bringing them into
alignment with the $\sigma_\circ^{opt}$ values.

The experimental situation presented above is far from ideal. While
the POR measurements give reliable estimates of the quasiparticle
scattering time from the resonance half-width, the spectrometer is
not calibrated to measure the absolute loss due to the sample.
Consequently, the absolute value of the conductivity is not obtained
from these measurements. Furthermore, it is usually the interlayer
conductivity that is measured, so one has to make the assumption
that the measured scattering time is isotropic. Nevertheless, the
obtained value of $\tau$ broadly agrees with other techniques such
as SdH and dHvA. In the case of the optical studies, neither
$\tau^{opt}$ or $\sigma_\circ^{opt}$ are obtained directly for the
three materials highlighted in this chapter, and the reported
$\tau^{opt}$ values are inconsistent with many other observations,
e.g. the dHvA and SdH effects. Nevertheless, it has been suggested
that both the Bechgaard and $\alpha$-phase BEDT-TTF salts exhibit
dramatic deviations from a simple Drude response on the basis of
these optical studies \cite{Schwartz98,DresselPRL03}. One of the
main pieces of evidence is the missing spectral weight in the
low-energy Drude peak. However, as already pointed out above, if one
takes the POR scattering time and the DC conductivity reported by
Bechgaard et al. \cite{Bech81}, then one finds that the Drude peak
in (TMTSF)$_2$ClO$_4$ accounts for $100\%$ of the free-carrier
spectral weight. Furthermore, the observation of magneto-oscillatory
phenomenon in many of these materials is hard to reconcile with the
notion of a Drude peak containing only $1\%$ of the free-carrier
spectral weight. Clearly, the only way to resolve these issues is
through the development of techniques which accurately measure the
low-energy electrodynamic properties (including DC) of these
materials. This ought to be a relatively straightforward task, but
available samples are often tiny and prone to twinning and
micro-cracking, which can severely influence such measurements.

Finally, we note that very extensive low-energy electrodynamic
measurements have been reported down to 0.1~cm$^{-1}$ for the
(TMTSF)$_2$PF$_6$ member of the Bechgaard family
\cite{Schwartz98,DresselPRB05,Donovan}, as have detailed AMRO
measurements \cite{Chash98}. Furthermore, the low-temperature
properties of this material are not influenced by anion ordering.
Therefore, it is highly desirable to make POR measurements on
(TMTSF)$_2$PF$_6$. However, this requires the application of
hydrostatic pressure ($>7$~kbar) in order to suppress a SDW phase
that occurs below 12~K under ambient pressure conditions. We note
that such techniques are currently under development.

\section{Summary and conclusions} \label{sec:12}

On the basis of detailed POR measurements, we present compelling
evidence that the low-energy magnetoelectrodynamics of three
contrasting organic conductors, (TMTSF)$_2$ClO$_4$,
$\alpha$-(BEDT-TTF)$_2$KHg(SCN)$_4$ and
$\kappa$-(BEDT-TTF)$_2$I$_3$, can be explained on the basis of a
conventional semiclassical Boltzmann theory, and that all three
materials exhibit coherent 3D band transport at liquid helium
temperatures. We demonstrate that there is nothing `magic' about the
angles at which DC resistance minima are observed in AMRO
experiments, findings that do not support the notion of a
fundamentally different thermodynamic ground states at, and away
from, the Lebed 'magic angles'. We also argue that the POR can
account for $100\%$ of the free-carrier spectral weight for
(TMTSF)$_2$ClO$_4$. Again, this finding appears to conflict with
claims of dramatic deviations from a simple Drude response on the
basis of broadband optical measurements. Finally, we propose that
the POR technique could be used to probe quasiparticles in nodal
superconductors.

\section{Acknowledgements}
This work was supported by the National Science Foundation
(DMR0239481) and by Research Corporation. The authors acknowledge
useful discussions with David Tanner, Victor Yakovenko and Martin
Dressel.

%


\printindex
\end{document}